\documentclass[preprint,onecolumn,showpacs,amsmath,amssymb,floatfix,pre]{revtex4}
\usepackage{amsfonts}
\usepackage{amssymb}
\usepackage{amsmath}
\usepackage{epsfig}
\begin{document}
\title{Driven lattice gas with nearest-neighbor exclusion: shear-like drive}
\author{Fabricio Q. Potiguar}
\email{potiguar@fisica.ufmg.br}
\author{Ronald Dickman}
\email{dickman@fisica.ufmg.br} \affiliation{Departamento de F\'\i
sica, ICEx, Universidade Federal de Minas Gerais, 30123-970, Belo
Horizonte, Minas Gerais, Brazil}

\begin{abstract}
We present Monte Carlo simulations of the lattice gas with
nearest-neighbor exclusion and Kawasaki (hopping) dynamics, under
the influence of a nonuniform drive, on the square lattice. The
drive, which favors motion along the +$x$ and inhibits motion in the
opposite direction, varies linearly with $y$, mimicking the velocity
profile of laminar flow between parallel plates with distinct
velocities. We study two drive configurations and associated
boundary conditions: (1) a linear drive profile, with rigid walls at
the layers with zero and maximum bias, and (2) a symmetric
(piecewise linear) profile with periodic boundaries. The transition
to a sublattice-ordered phase occurs at a density of about $0.298$,
lower than in equilibrium ($\rho_c \simeq 0.37$), but somewhat
higher than in the uniformly driven case at maximal bias ($\rho_c
\simeq 0.272$). For smaller global densities ($\rho \leq 0.33$),
particles tend to accumulate in the low-drive region.  Above this
density we observe a surprising reversal in the density profile,
with particles migrating to the high-drive region and forming
structures similar to force chains in granular systems.
\end{abstract}

\maketitle

\section{Introduction}

An important model of interacting particles in statistical mechanics
is the lattice gas. In this model the particles have
nearest-neighbor interactions which can be either attractive
(ferromagnetic) or repulsive (antiferromagnetic). In equilibrium,
such systems have been used extensively as models of simple fluids.
Nonequilibrium versions of this model have also been widely studied
\cite{sch-livro,marro}, generally in the form of a lattice gas with
periodic boundaries under a constant drive that biases the hopping
rates along one of the principal axes of the lattice \cite{kat83}.
Lattice gases with biased hopping are also known as {\em driven
diffusive systems} (DDS) \cite{sch-livro}; the repulsive version
\cite{leu89,dic90} serves as a model for fast ionic conductors
\cite{boy79}.

The lattice gas with nearest-neighbor exclusion (NNE) is the
infinite repulsion limit of the ordinary repulsive lattice gas.
Here, particles are forbidden to occupy the same or neighboring
sites; the minimum allowed interparticle separation is that for
second neighbors. Hard-core exclusion is the only interaction
between the particles. (This model also represents the zero-neighbor
limit of the Biroli-M\'ezard lattice glass model \cite{bir02}.) The
equilibrium version was studied in \cite{gau65,ree66,run66,gau67,baxter}
both theoretically and numerically, for various lattice types. These
analyses show that the model exhibits a continuous phase transition
to an ordered state at a critical density $\rho_c$ ($\rho_c
\approx0.37$ on the square lattice), with exponents that belong to
the Ising universality class.

More recently, nonequilibrium versions of the NNE model were studied
\cite{dic01,szo02}. In Ref. \cite{dic01}
(using nearest-neighbor hopping dynamics on the square lattice),
it was found that the
critical density varies with drive intensity: the higher the drive,
the lower the critical density. The transition is continuous for low
bias but becomes first order if the bias strength is $\geq 0.75$.
Above the transition density, the system separates into regions of
low and high local density, with the high-density region essentially
frozen. Szolnoki and Szab\'o \cite{szo02} extended the dynamics to
include next-nearest-neighbor (diagonal) hops, and observed a
similar variation of the critical density with drive strength, but
with a homogeneous stationary state.
Continuous phase transitions in
this version of the model fall in the Ising class, as the
equilibrium case.

Here we consider a hard-core DDS with nearest-neighbor hopping
dynamics, in which the drive is {\it nonuniform}. The simplest case
is that in which the bias, along the $x$ direction, varies linearly
along the perpendicular ($y$) direction. The motivation for this
drive configuration is to model a system of hard particles in a
uniformly sheared fluid, or a granular system under shear.
Specifically, the probability for a particle at $y$ to attempt a
jump to the right ($x \to x+1$) is given by:
\begin{equation}
\label{pro-r}
P_r(y)=\frac{1}{4}\left(1+\frac{y-1}{L-1}\right),
\end{equation}
for $y=1$, 2,...,$L$, on a square lattice of $L^2$ sites. The
corresponding probability for attempting to hop to the left is
$P_l(y)=1/2-P_r(y)$, while hops in the $\pm y$ directions are
attempted with probabilities of 1/4. From Eq. (\ref{pro-r}) it is
clear that a particle at $y=1$ experiences no bias, while one at
$y=L$ cannot jump to the left. We present numerical results from
Monte Carlo (MC) simulations of the model. One point to be noted is
that the drive cannot overcome the repulsion between particles:
attempted moves are accepted if and only if the nearest-neighbor
exclusion condition is respected.

Our main objective is to obtain the phase diagram and critical
properties of the model. We determine the critical density and study
the behavior of the order parameter and the stationary current as
functions of density.  Of particular interest are the current and
the density profiles as functions of $y$. Our simulations indicate
that at low densities, particles tend to accumulate in the low-drive
region, and that as we increase the density, they migrate to the
high-drive region. This is surprising since, in the uniform-drive
case \cite{dic01}, a large bias is required to cause particle
aggregation. The density and current profiles change dramatically at
a density of about $0.33$. In the following section (II) we detail
the model and simulation procedure.  In Section III, we present
numerical results and discussions. Final considerations are reserved
for section IV.

\section{Simulations}

We consider a square lattice of length $L$ ($L^2$ sites) with $N$
particles ($N<L^2/2$).  (Most of our studies use $L=100$.) The
initial configuration is prepared via random sequential adsorption
(RSA) \cite{mea87,dic91} of particles, always respecting the
excluded-volume condition.  Once all of the $N$ particles have been
inserted, the biased hopping dynamics begins. A particle is selected
at random and assigned a new (trial) position at one of the nearest
neighbor sites, with probabilities as discussed above.  If the trial
position does not violate the exclusion constraint, the move is
accepted, otherwise it is rejected. Each MC time unit corresponds to
$N$ attempted moves. We follow the evolution for a total of $10^6$
MC time units. We use 20 values of density $\rho = N/L^2$ to
investigate the transition, ranging from $\rho = 0.1$ to $\rho=
0.37$.

We study two kinds of drive configuration and associated boundary
conditions. The first, employed with the linear bias profile, Eq.
(\ref{pro-r}), imposes rigid walls at $y=1$ and $y=L$ (moves to
$y=0$ or $y=L+1$ are prohibited). The second one involves a drive
profile symmetric about the mid-plane ($y=L/2$), such that the bias
is the same at $y=1$ and $y=L$.  In this case we apply {\em
periodic} boundary conditions in the $y$-direction. The symmetric
drive profile used in the case of periodic boundaries is piecewise
linear:
\begin{equation}
\label{sim-pro}
P_r(y)=\left\{
\begin{array}{ll}
\frac{1}{4}\left(1+\frac{y-1}{L/2-1}\right), & \text{if $y\leq L/2$}\\
\frac{1}{4}\left(1+\frac{L-y}{L/2}\right)~, & \text{otherwise}
\end{array}
\right.
\end{equation}
This equation guarantees that particles crossing the boundary in the
$y$ direction are subject to the same bias. We use periodic
boundaries in the $x$-direction (parallel to the drive) in both
cases.

\section{Results}

\subsection{Stationary global properties}

To characterize the behavior of the model we study the order
parameter and the stationary current. The order parameter is defined
as the difference in sublattice occupancies per particle,
\begin{equation}
\label{ord-par}
\phi=\frac{N_A-N_B}{N},
\end{equation}
where $N_{A(B)}$ is the number of particles in sublattice $A(B)$.
The current is defined as the difference between the number of jumps
along the drive less the number contrary to it, per site and unit
time.

\begin{figure}[!tbp]
\rotatebox{0}{\epsfig{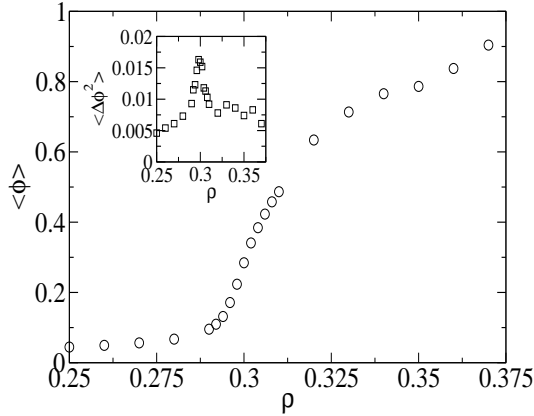}}
\caption{Order parameter (circles)
versus overall density, $L=100$. The inset shows its variance. The
peak in $\left<\Delta\phi^2\right>$ is around $\rho=0.298$.} \label{OP}
\end{figure}

In Fig. \ref{OP} we show the average value of the order parameter
and, in the inset, its corresponding fluctuation. The transition point is
characterized by a sudden rise in the order parameter and by a peak
in its variance.  The data suggest a continuous transition to
sublattice ordering at a critical density of $\rho_c=0.298$.  The
critical density is lower than that for the equilibrium transition
\cite{ree66} of $\rho_c \approx 0.37$. In the {\em uniformly} driven
system studied in \cite{dic01}, the transition (for maximal bias,
for $L=100$) occurs at a density of 0.272, and is {\it
discontinuous}. A transition density of $\rho_c =0.30$ in the
uniformly driven system corresponds to a bias of 0.75 ($P_r = 3/8$);
at this point the transition is discontinuous. (The transition in
the uniformly driven system is continuous for $\rho \leq 0.6$.) The
apparently continuous transition in the presence of a nonuniform
drive is likely to be due the fact that the system does not order
all at once. We will return to this point later.
\vspace{2em}

\begin{figure}[h]
\rotatebox{0}{\epsfig{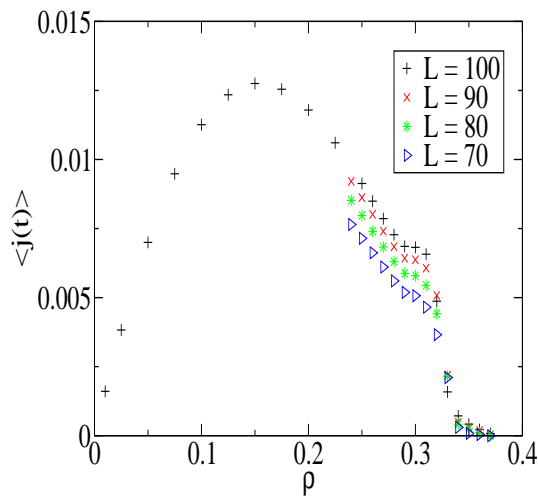}}
\caption{Average stationary current versus density for $L=100$, $90$,
$80$ and $70$.}
\label{cur}
\end{figure}

In Fig. \ref{cur}, we show the stationary current, averaged over the region
$16 < y < L-16$, for each lattice length $L$, to avoid strong wall effects.
This quantity displays the same behavior as in the uniformly driven case:
it increases at small densities (reflecting the increasing number of
carriers) and decreases for larger densities (due to the reduction
in available space for movement). The maximum value of $\langle
j(t)\rangle$ falls at roughly in the same density as in the uniform
drive case \cite{dic01}. Interestingly, the phase transition near $\rho=0.298$
is associated with a plateau in the current.

The order parameter and the stationary current present strong
fluctuations for densities above $\rho=0.32$. The evolution of these
quantities typically displays sudden jumps between the ordered and
the disordered state, a fact already observed in the uniformly
driven case. The drive provokes formation of organized structures
while their thermal motion provides a mechanism for breaking such
clusters. We will see that such behavior may find a parallel in
actual physical systems.

\begin{figure}[!tbp]
\rotatebox{0}{\epsfig{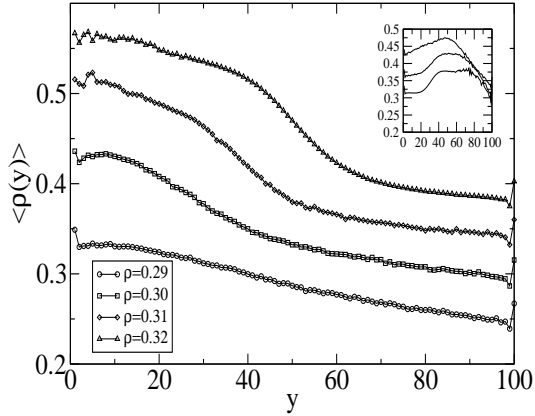}}
\caption{Stationary density profile $\rho(y)$ for various global
densities. The curves for the three denser states are shifted
upwards by $0.15$, $0.1$ and $0.05$, respectively. Inset: same
quantity for densities $0.33$, $0.34$, and $0.35$ (from top to
bottom). The first two curves are shifted upwards by $0.1$ and
$0.05$.} \label{RhoY}
\end{figure}

\begin{figure}[!tbp]
\rotatebox{0}{\epsfig{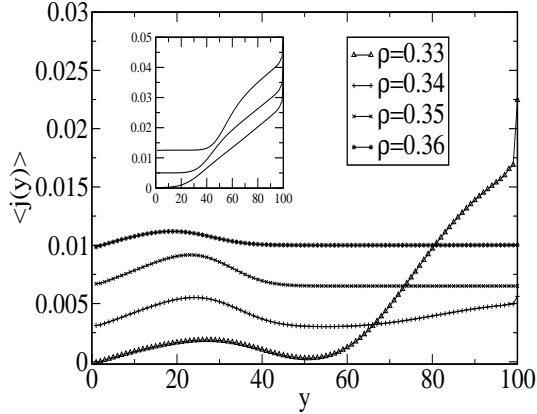}}
\caption{Stationary current profile $j(y)$ for global densities as
indicated. The data for $\rho=0.36$, $0.35$ and $0.34$ data are
shifted upwards by $0.01$, $0.007$ and $0.003$ respectively. Inset:
average stationary current for densities (top to bottom): $0.32$,
$0.31$ and $0.30$. The curves is shifted upwards by $0.0125$, the
second by $0.005$.} \label{CurY}
\end{figure}

\subsection{Density and current profiles}

Of interest is how the system organizes as the density increases. To
understand this, we determine the stationary density and current
profiles, $\rho(y)$ and $j(y)$, respectively. These quantities are
shown in Figs. \ref{RhoY} and \ref{CurY}. The immediate conclusion
that can be drawn from these plots is that particles concentrate in
the low-drive region (lower half of the lattice), for densities up
to $\rho=0.33$.

Enhanced particle concentration in the low-drive region is in fact
observed for densities as low as $\rho=0.20$. This is surprising
given the finding that, in a uniformly driven system, a strong drive
favors order, i.e., provokes a reduction in the critical density. On
this basis one might expect the high-drive region to order first
(i.e., at a lower global density), which would require particles to
concentrate in the {\it high}-drive region.  In fact just the
opposite occurs: for a global densities $\rho \leq 0.32$, the
density profile $\rho(y)$ (Fig. \ref{RhoY}) is highly skewed to the
region around $y=0$, where the bias is small.  Note  that the local
density is $\geq 0.37$ in this region (for $\rho$ between $0.30$ and
$0.32$), that is, greater than or equal to $\rho_c$ in equilibrium.
The density profile decays monotonically with increasing $y$ (except
for small density oscillations induced by the wall at $y=0$). For
$\rho=0.33$, the local density instead increases with $y$, reaching
a peak near $y=47$, after which it decays in an approximately linear
fashion until $y=L$.  For a certain range of global densities
$\rho$, the local density near $y=1$ actually {\it decreases} with
increasing $\rho$. Note also that, as may be seen from Fig.
\ref{CurY}, the current density in the high-drive region is zero for
$\rho \geq 0.35$.

\subsection{Ordering and jamming}

Ordering occurs first in the low-drive region. The high density in
this region is also reflected in the current profile (Fig.
\ref{CurY}). The current is much smaller in the small-$y$ region
(low-bias) and increases monotonically with $y$.  The order
parameter profile (not plotted here), follows the same pattern as
the density profile. It clearly shows that ordering occurs in the
low-drive region for densities between $0.30$ and 0.32, as described
above.

A possible explanation for the surprising reversal of the density
and current profiles with increasing global density is related to
the formation of a jammed region, as observed in the uniformly
driven system \cite{dic01}. When the global density is too low for
such a region to form, particles tend to collect in the low-drive
region because a strong drive tends to destroy the local
correlations needed for particles to pack to high density, even if
such packing does not result in long-range order.
The depletion of the high-drive region appears to be the reason for the
plateau in the current observed around $\rho =0.30$ in Fig. \ref{cur}.
Fig. \ref{RhoY} shows that as the global density increases, the local density
in the high-drive region remains nearly constant, so the current hardly
varies.

When, on the other
hand, the global density is sufficiently high for a jammed region to
form, it appears in the high-drive region, leading to an
irreversible accumulation of particles there, so that the low-drive
region has fewer particles than at lower global densities, for which
there is no jammed region.

\begin{figure}[!tbp]
\rotatebox{0}{\epsfig{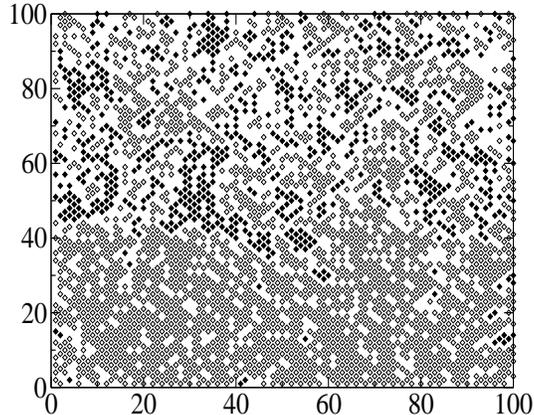}}
\caption{Particle configuration at density $\rho=0.31$, after
$10^{6}$ MC steps (system size $L=100$). Open and filled symbols
represent particles in different sublattices. The drive is directed
to the right and increases in the vertical direction.}
\label{f-conf-1}
\end{figure}

To illustrate these ideas, we show in Fig. \ref{f-conf-1} a
configuration for $\rho=0.31$ and $L=100$. As expected, the
low-drive region is very dense and contains few mobile particles. In
the uniformly driven system (at maximum bias) one observes, at this
density, formation of ``herringbone" pattern of diagonal stripes,
pointing along the drive, with particles in this structure
essentially frozen.  In the present case the low-drive region is
highly ordered, with almost all particles in he same sublattice, but
there is no sign of the herringbone pattern. The high-drive region
is disordered, permitting the high currents and lower densities
reported above.  Several clusters of particles exist in the
high-drive region, but they are not large enough to cause jamming.

We may now identify two factors leading to the continuous variation
of the order parameter with density shown in Fig. 1.  One reason is
that ordering begins in the low-bias region.  Studies of the
uniformly driven system show that the transition is continuous under
a weak bias. The second reason is that the width of the ordered
region grows continuously with increasing density.

For global densities above about $0.34$, the situation, as noted, is
completely changed. Particles now accumulate in the high-drive
region, and the density profile exhibits a maximum in the central
region (intermediate drive strength).  At these higher densities the
current profile displays a peak in the low-drive region.  The peak
shifts to smaller $y$ (smaller drive intensity) and decreases in
amplitude as the global density is increased.

\begin{figure}[!tbp]
\rotatebox{0}{\epsfig{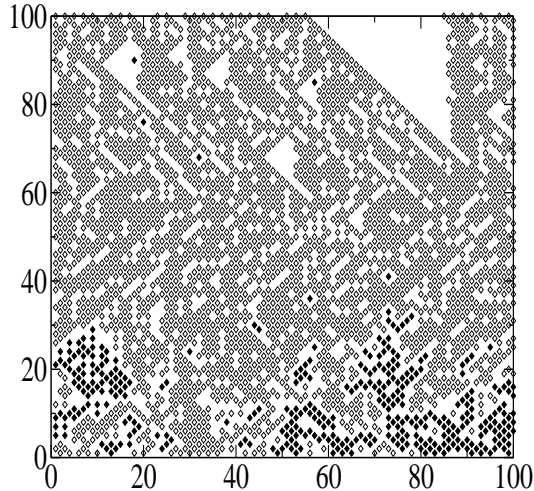}}
\caption{Particle configuration at density $\rho=0.35$, after $10^6$
steps ($L=100$).} \label{f-conf-2}
\end{figure}

Fig. \ref{f-conf-2} shows a typical configuration at global density
$\rho=0.35$.  Evidently the long diagonal line of particles at the
upper right is associated with jamming in the high-drive region. The
empty triangular region implies a decrease in the local density with
increasing $y$. Particles are not free to enter this region since
all particles along the diagonal edge are jammed. These observations
are supported by the the density and current profiles (Figs. 3 and
4). The density is roughly constant in the middle portion of the
lattice and begins to decrease near $y=78$, where the empty
triangular region begins. The current is only appreciably different
from zero in the lower portion of the lattice, as signalled in Fig.
\ref{f-conf-2} by the presence of particles both sublattices. The
diagonal edges observed in configurations at this density (always in
the high-drive region), are extremely long-lived structures, since
only the particle at the tip of the line can move without violating
the exclusion constraint. The particles at the tips of the diagonal
are blocked by others, giving a virtually infinite lifetime to this
structure.

\begin{figure}[!tbp]
\rotatebox{0}{\epsfig{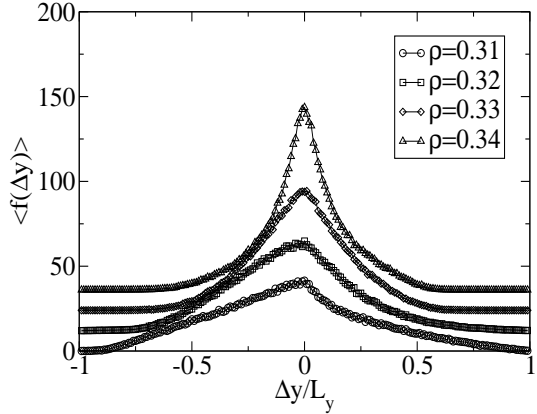}}
\caption{Histogram of displacements $\Delta y$ during the
thermalization period ($L=100$). The curves for the three denser
states are shifted upwards by $36$, $24$ and $12$, respectively.}
\label{ydisp-100}
\end{figure}

Of note are the large number of voids and diagonal strings of
particles in the high-bias region of Fig. \ref{f-conf-2}. This hints
at the possibility that the entire jammed structure is induced by
the drive  in the early stages of the evolution. To verify this
conjecture, we show in Fig. \ref{ydisp-100} the distribution of
$y$-displacements (averaged over all particles), during
thermalization, for four different global densities. The slightly
asymmetric curves are a signature of the accumulation of particles
in the low-bias region; they have a higher probability to migrate
downwards than upwards. For densities below $0.31$, and above
$0.34$, no asymmetry can be detected in the curves for $f(\Delta
y)$. The distribution narrows sharply with increasing global
density, so that for most particles only small $y$-displacements are
possible.

\begin{figure}[!tbp]
\rotatebox{0}{\epsfig{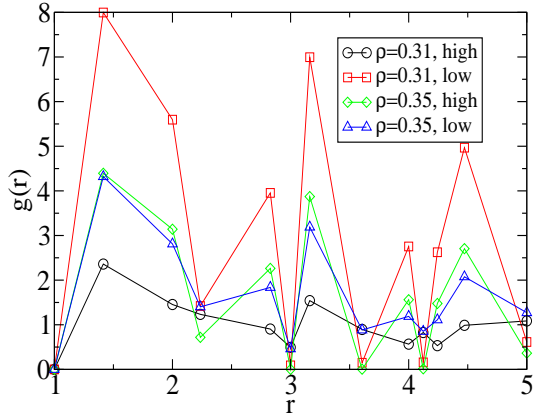}}
\caption{Pair distribution function for the high- and low-drive
regions, for densities $\rho=0.31$ and $\rho=0.35$.} \label{g-of-r}
\end{figure}

To study correlations between the particles, we determine the radial
distribution function, $g(r)$ in the high- and low-drive regions
(Fig. \ref{g-of-r}). This function is proportional to the
probability of finding a pair of particles separated by a distance
$r$, and is normalized so that $g \to 1$ as $r \to \infty$. For
purposes of determining $g(r)$, the low-drive region is taken as the
strip $6 \leq y \leq 14$, while the high-drive region comprises $86
\leq y \leq 94$.

The $g(r)$ curves for global density $\rho=0.31$ show that the high-
and low-drive structures are markedly different. In the low-bias
region the peaks are much larger due to the sublattice ordering
associated with packing of particles (as evidenced by the
configuration in Fig. \ref{f-conf-1}), and is compatible with the
existence of long-range order. The high-drive region shows little
structure; the oscillations in $g(r)$ decay rapidly with distance.
The picture for $\rho = 0.35$ is quite different. The sharpness of
the peaks in the curve for $\rho=0.35$ in the high-bias region
reflects the very different sublattice densities, as does the fact
that $g \simeq 0$ for $r=3$, $\sqrt{13}$ and $\sqrt{17}$. The radial
distribution function in the low-drive region, for this density,
exhibits less structure, indicating the more equal sublattice
occupancies.

\subsection{Boundary effects}

The density profiles of Fig. \ref{RhoY} show that particles tend to
accumulate in the layers in contact with the rigid walls ($y=1$ and
$y=L$). This accumulation is due to excluded volume effects
\cite{dic97} that entropically favor enhanced densities at rigid
surfaces.  (A certain fraction of the volume excluded by the
particles in these layers overlaps with the volume excluded by the
wall, thereby leaving more space for the remaining particles.) The
effect of this accumulation shows up in the current profile (Fig.
\ref{CurY}), as a sharp peak at $y=L$. Although the number of
particles is larger in these layers, it is not yet enough to provoke
a jammed structure, therefore allowing a high current. At higher
global densities, the number of particles is large enough for them
to block the way and there is no current at all near the boundary at
$y=L$.

\begin{figure}[h]
\rotatebox{0}{\epsfig{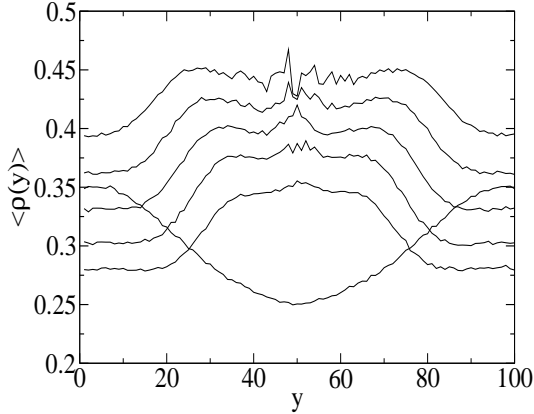}}
\caption{Stationary density profile under the symmetric drive
profile, Eq. (\ref{sim-pro}). The curve with a central minimum
corresponds to $\rho=0.30$. The others are for densities (from
bottom to top): $\rho=0.31,0.32,0.33,0.34,0.35$. The last five
curves were shifted upwards by: $0.1$, $0.08$, $0.06$, $0.04$ and
$0.02$, respectively.} \label{RhoY-SD}
\end{figure}

\begin{figure}[h]
\rotatebox{0}{\epsfig{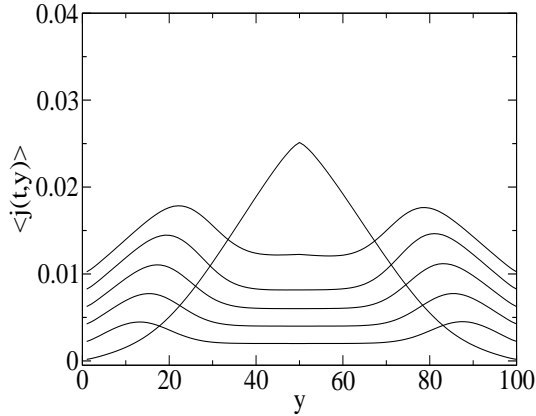}}
\caption{Average stationary current profile for a system subjected
to the drive of Eq. (\ref{sim-pro}). The single peaked curve is for
$\rho=0.30$. The others are for the same densities as in fig.
\ref{RhoY-SD} (from top to bottom). The curves for $\rho=0.31$ to
$0.35$ were shifted by $0.01$, $0.008$, $0.006$, $0.004$ and
$0.002$.} \label{CurY-SD}
\end{figure}

The question naturally arises as to whether the surprising changes
in the density and current profiles (as functions of global
density), described above, are somehow induced by the rigid walls at
$y=1$ and $y=L$. To see that this is not the case, we show in Figs.
\ref{RhoY-SD} and \ref{CurY-SD} the density and current profiles in
the system with a symmetric drive profile, Eq. (\ref{sim-pro}), and
periodic boundaries in the $y$ direction. We find that for $\rho
\leq 0.30$, particles are concentrated the low-drive region, just as
in the system with rigid walls. Above this density the pattern
changes to one in which they concentrate in the high-drive region.
As we increase the number of particles, they continue to accumulate
in this region, and the local density in the low-drive region is
smaller than observed at a global density of $0.30$ (up to
$\rho=0.34$). Interestingly, around the position of maximum bias,
the local density decreases in a region of approximately 15 lattice
spacings. This suggests the presence of diagonal voids that prevent
homogenization of the lattice. These structures become more evident
with increasing density, but not as dramatically as in the presence
of walls.

The behavior of the current parallels these trends, just as in the
studies with rigid walls. For densities up to $\rho=0.30$ the
current profiles display a peak in the high-drive region. Increasing
the value of $\rho$ beyond this point dramatically changes the
profile. The peaks are now located in the region of intermediate
drive strength, and the current in the maximum-bias region falls to
zero. As before, the peaks decrease in amplitude and shift towards
the low-drive region as we increase the density. The same
explanation as before holds: as the density is increased more and
more particles accumulate near the central region, therefore
narrowing the window where particles are mobile (while decreasing
the effective number of mobile carriers).  Thus rigid wall are not
the cause of the reversals observed in the density and current
profiles.

\section{Conclusions}

We studied a lattice gas with nearest-neighbor exclusion driven by a
nonuniform, shear-like drive, on the square lattice, under
nearest-neighbor hopping dynamics.
The problem is of interest both as an example of the surprising behavior
to be found in a simple nonequilibrium system, and as a toy model for
a granular or colloidal system under shear.
We find that the model undergoes
a continuous order-disorder transition at a critical density of
about $\rho_c=0.298$. This is unlike the uniformly driven model, in
which the transition is discontinuous for a bias $\geq 0.75$. The
stationary current follows roughly the same trends as in the
uniformly driven case, but exhibits a plateau in the neighborhood of
the phase transition.

Our results show that this transition is due to the concentration of
particles at the low-bias region, for global densities between
$\rho=0.30$ and $0.32$.  Remarkably, the nonuniform drive induces a
highly nonuniform density profile, expelling particles from the
high-bias region.  The effect is sufficiently strong to induce
sublattice ordering in the low-bias region.  Thus the drive favors a
class of configurations that, on the basis of entropy maximization,
are extremely unlikely.  Note that at these densities there is no
jamming, i.e., the system is ergodic.  Migration of particles to the
low-bias region appears to derive from the destruction of
short-range correlations (required for efficient packing), by the
drive.

For higher densities, we observe a completely inverted picture, with
formation of jammed structures in the high-drive region, while
particles outside this region remain mobile. The jammed region is
characterized by a dense ($\rho \geq 0.37$) strip of particles; at
higher global densities this region displays long diagonal chains of
particles associated with voids.

Preliminary studies show that the above-mentioned effects also
appear if half the system has {\it no bias} while the remainder is
subject to maximum bias \cite{fut01}.

The jammed diagonal structures appearing at high density may be seen
as an instance of force chains in granular systems \cite{cat98}. By
applying a shear drive to a packing of grains, interparticle
contacts are built, leading to the formation of chain structures
that carry the interparticle forces, which sustain further stress in
that direction. In experiments, the system yields under a
sufficiently large stress, breaking the chains. Since in our model
the interparticle interactions are infinite, so is the yield stress.
Therefore, as long as the drive is applied, we observe such ideal
force chains. For a system in which the density varies while the
drive is on, such force chains can arise at much lower densities
\cite{fut01}.

A thermal system composed of particles that have highly repulsive
interactions, albeit not infinite ones, is a colloidal suspension.
In fact, the experiments of Bertrand {\em et al.} \cite{ber02} show
a similar effect to the slow relaxation observed here. They studied
suspensions at several densities under the influence of shear. The
suspension presents shear-thickening at intermediate densities,
where a small vibration drives the system back to a fluid state. At
higher densities, after shear is applied, the suspension forms a
paste, becoming trapped in this jammed state. In our case, thermal
motion, or vibration, is always present, as well as shear, so we do
not observe the breakdown of the jammed state at intermediary
densities, but a slow relaxation towards a denser state. In a
granular system, which is, by definition, athermal, effects
analogous to those of thermal agitation can be produced by shaking.
This raises the possibility that the behavior identified in the
sheared lattice gas might also be observed in a sheared packing if,
besides the shear drive, continuous shaking were applied to the
grains. This suggests that the model studied here can be extended to
study the dynamics of certain complex fluids, a subject we intend to
explore in future work.

\begin{acknowledgments}
We thank the Brazilian agencies CAPES, CNPq and Fapemig for support.
\end{acknowledgments}

\bibliographystyle{apsrev}

\begin{thebibliography}{100}
\bibitem{fut01} Studies of a system with half zero and half maximum drive, and
a driven system
with variable density are in progress.
\end{thebibliography}

\end{document}